\documentclass[conference,a4paper]{IEEEtran}
\usepackage{xcolor}
\usepackage{balance}
\usepackage{amsmath, amsfonts, amssymb}
\usepackage{tikz}
\usepackage[caption=false,font=footnotesize]{subfig}

\usepackage{graphicx}
\ifCLASSOPTIONcompsoc
 \usepackage[caption=false,font=normalsize,labelfont=sf,textfont=sf]{subfig}
\else
 \usepackage[caption=false,font=footnotesize]{subfig}
\fi
\begin{document}
%
\title{A Beam-Switched Sparse Circular Array \\for Passive Wi-Fi Sensing}

\author{\IEEEauthorblockN{
Yang Chen\IEEEauthorrefmark{1}\IEEEauthorrefmark{2},  
Riccardo Bersan\IEEEauthorrefmark{1},   
Daniele Piazza\IEEEauthorrefmark{1},    
Stefano Savazzi\IEEEauthorrefmark{3},    
Michele D'Amico\IEEEauthorrefmark{2}      
}                                     
\IEEEauthorblockA{\IEEEauthorrefmark{1}
Adant S.r.l., Padova, Italy}
\IEEEauthorblockA{\IEEEauthorrefmark{2}
Politecnico di Milano, DEIB, Milan, Italy, yang.chen@polimi.it}
\IEEEauthorblockA{\IEEEauthorrefmark{3}
Consiglio Nazionale delle Ricerche (CNR), IEIIT, Milan, Italy,
stefano.savazzi@cnr.it}

}



\maketitle

\begin{abstract}
Wi-Fi sensing applications require unambiguous angle-of-arrival (AoA) estimation over the full azimuth plane. Uniform circular arrays (UCAs) are shown to remove this ambiguity, but deploying the four RF chains typically available on commercial Wi-Fi chipsets into a sparse layout suffers from spatial undersampling. Such undersampling produces strong grating lobes. To overcome this hardware bottleneck, we first establish a generalized array manifold incorporating element radiation patterns, revealing the suitability and degradation mechanisms of directional UCAs. Then we propose an angularly staggered beam-switched UCA architecture that leverages the spatial complementarity. Numerical evaluations demonstrate that the proposed scheme effectively mitigates grating-lobe ambiguities without increasing the number of RF chains, significantly reducing AoA estimation outliers and improving omnidirectional tracking performance.
\end{abstract}

\vskip0.5\baselineskip
\begin{IEEEkeywords}
 Wi-Fi sensing, switched antenna arrays, sparse arrays, AoA estimation, grating lobes.
\end{IEEEkeywords}

%

\section{Introduction}

With the advancement of IEEE 802.11bf standardization~\cite{802.11}, passive sensing based on Wi-Fi Channel State Information (CSI) has emerged as a prominent paradigm in Integrated Sensing and Communication (ISAC)~\cite{WiDFS,Wi-Nod}. A prerequisite for ubiquitous indoor sensing is the provision of unambiguous angle-of-arrival (AoA) estimation across the full azimuth plane. However, most existing algorithms rely on the assumption of half-wavelength uniform linear arrays (ULAs)~\cite{WiDFS,SpotFi}. This not only suffers from $180^\circ$ front-back mirror ambiguity, but also overlooks the requirement of communication MIMO for sparse arrays to achieve spatial diversity~\cite{TGnChannel}. Uniform circular arrays (UCAs) can typically eliminate this front-back mirror ambiguity~\cite{DeepAoA}. However, in commercial access points (AP) constrained to only four RF chains, the deployment of a sparse UCA enlarges the aperture, but severe spatial undersampling leads to strong grating lobes~\cite{Mishra2024}.

To address the grating lobe issue caused by limited channels and sparse elements, existing research generally follow three routes: First, using high-speed RF switches to poll antennas for coherent snapshot stitching~\cite{Morano2024,Simoncic2025}. However, this method relies heavily on static assumptions and is easily disrupted by Doppler shifts and target displacement. Second, employing virtual array synthesis~\cite{15radar,Zhu2023} or subarray optimization~\cite{Mishra2024}. However, active orthogonal waveforms~\cite{15radar} are difficult to reconcile with passive Wi-Fi sensing, coarse prior estimation~\cite{Zhu2023} requires reliable initial angles. Deep-learning-based subarray planning~\cite{Mishra2024} relies on redundant channels to build priors and suffers from limited generalization. Third, incorporating kinematics or spatio-temporal filtering to assist in disambiguation~\cite{Zhu2025}, but such high-layer post-processing algorithms still depend on reliable spatial decisions provided by the underlying layer. In addition, these methods generally assume ideal isotropic elements, ignoring the directional radiation characteristics caused by reflections from internal heat sinks and ground planes~\cite{Jackson2015,Biguesh2009}.

To this end, we first establish a generalized array manifold model integrating array topology and radiation patterns. Manifold analysis shows that while antenna directivity further compresses the effective field of view (FoV) of ULAs, it serves as a spatial filter for UCAs to suppress back grating lobes. However, sparse 4-UCAs still face sidelobe ambiguity at transition angles. Then we propose an angularly staggered beam-switched UCA architecture, which directly suppresses grating lobes through joint consensus spectrum synthesis using only two complementary 4-UCA bases, without coherent snapshot stitching or complex matrix completion. Numerical results demonstrate that the proposed four-RF-chain architecture achieves AoA estimation performance approaching that of an eight-channel omnidirectional reference array, while substantially reducing angular ambiguities and tracking outliers compared with a conventional four-element circular array.

\section{Manifold Modeling and Directivity Analysis}
Consider a typical indoor Wi-Fi sensing scenario, where a commercial AP is equipped with $M=4$ physical RF receive chains. In the 5~GHz band, a sounding frame received in OFDM communication contains $K$ orthogonal subcarriers with frequencies $f_k$ ($k=1,\dots,K$). Within a single-frame CSI packet snapshot, considering all $M$ RF channels, the CSI vector $\mathbf{h}_k \in \mathbb{C}^{M \times 1}$ received on the $k$-th subcarrier can be expressed as a coherent multipath superposition model~\cite{Zhang2021Overview}:
\begin{equation}\label{equ:1}
\mathbf{h}_k = \sum_{l=1}^{L} \gamma_l e^{j \phi_{D, l}} e^{-j 2\pi f_k \tau_l} \mathbf{a}(\theta_l),
\end{equation}
where $L$ is the total number of multipath components received. For the $l$-th path, $\gamma_l$, $\phi_{D, l}$, and $\tau_l$ represent the complex attenuation, instantaneous Doppler phase shift, and reference propagation delay arriving at the array origin, respectively. $\mathbf{a}(\theta_l) \in \mathbb{C}^{M \times 1}$ is the array manifold vector, incorporating both the AoA parameter $\theta_l$ and the antenna radiation patterns. Since all RF channels share the same local oscillator within a single snapshot, the macroscopic propagation delay $e^{-j 2\pi f_k \tau_l}$ and the Doppler phase $e^{j \phi_{D, l}}$ are identical in all antennas. Therefore, this section focuses on the spatial steering vector influenced by the antenna array distribution and the element radiation patterns, i.e. the array manifold $\mathbf{a}(\theta)$, and its underlying physical impact mechanisms.

\subsection{Generalized Array Manifold}

\newsavebox{\figonea}

\savebox{\figonea}{%

\tikzset{every picture/.style={line width=0.75pt}} 

\begin{tikzpicture}[x=0.75pt,y=0.75pt,yscale=-1,xscale=1]

\draw  (83.62,186.28) -- (365.68,186.28)(180.24,74.5) -- (180.24,287.28) (358.68,181.28) -- (365.68,186.28) -- (358.68,191.28) (175.24,81.5) -- (180.24,74.5) -- (185.24,81.5)  ;
\draw  [color={rgb, 255:red, 74; green, 144; blue, 226 }  ,draw opacity=1 ][dash pattern={on 4.5pt off 4.5pt}] (108.42,186.28) .. controls (108.42,146.77) and (140.45,114.74) .. (179.96,114.74) .. controls (219.47,114.74) and (251.5,146.77) .. (251.5,186.28) .. controls (251.5,225.79) and (219.47,257.82) .. (179.96,257.82) .. controls (140.45,257.82) and (108.42,225.79) .. (108.42,186.28) -- cycle ;
\draw  [color={rgb, 255:red, 74; green, 144; blue, 226 }  ,draw opacity=1 ] (180.24,115.02) -- (251.5,186.28) -- (180.24,257.53) -- (108.98,186.28) -- cycle ;
\draw  [fill={rgb, 255:red, 74; green, 144; blue, 226 }  ,fill opacity=1 ] (103.82,186.28) .. controls (103.82,183.43) and (106.13,181.11) .. (108.98,181.11) .. controls (111.83,181.11) and (114.15,183.43) .. (114.15,186.28) .. controls (114.15,189.13) and (111.83,191.44) .. (108.98,191.44) .. controls (106.13,191.44) and (103.82,189.13) .. (103.82,186.28) -- cycle ;
\draw  [fill={rgb, 255:red, 74; green, 144; blue, 226 }  ,fill opacity=1 ] (175.08,257.53) .. controls (175.08,254.68) and (177.39,252.37) .. (180.24,252.37) .. controls (183.09,252.37) and (185.4,254.68) .. (185.4,257.53) .. controls (185.4,260.39) and (183.09,262.7) .. (180.24,262.7) .. controls (177.39,262.7) and (175.08,260.39) .. (175.08,257.53) -- cycle ;
\draw  [fill={rgb, 255:red, 74; green, 144; blue, 226 }  ,fill opacity=1 ] (246.33,186.28) .. controls (246.33,183.43) and (248.64,181.11) .. (251.5,181.11) .. controls (254.35,181.11) and (256.66,183.43) .. (256.66,186.28) .. controls (256.66,189.13) and (254.35,191.44) .. (251.5,191.44) .. controls (248.64,191.44) and (246.33,189.13) .. (246.33,186.28) -- cycle ;
\draw  [fill={rgb, 255:red, 74; green, 144; blue, 226 }  ,fill opacity=1 ] (175.08,115.02) .. controls (175.08,112.17) and (177.39,109.86) .. (180.24,109.86) .. controls (183.09,109.86) and (185.4,112.17) .. (185.4,115.02) .. controls (185.4,117.87) and (183.09,120.18) .. (180.24,120.18) .. controls (177.39,120.18) and (175.08,117.87) .. (175.08,115.02) -- cycle ;
\draw [line width=1.5]    (302,115.5) -- (187.23,179.82) ;
\draw [shift={(183.75,181.77)}, rotate = 330.73] [fill={rgb, 255:red, 0; green, 0; blue, 0 }  ][line width=0.08]  [draw opacity=0] (13.4,-6.43) -- (0,0) -- (13.4,6.44) -- (8.9,0) -- cycle    ;
\draw    (289.8,184.59) .. controls (297.65,166.04) and (291.94,150.2) .. (280.4,135.78) ;
\draw [shift={(278.53,133.52)}, rotate = 49.64] [fill={rgb, 255:red, 0; green, 0; blue, 0 }  ][line width=0.08]  [draw opacity=0] (8.93,-4.29) -- (0,0) -- (8.93,4.29) -- cycle    ;

\draw (369.67,181.33) node [anchor=north west][inner sep=0.75pt]  [font=\normalsize] [align=left] {$\displaystyle x$};
\draw (188.67,67.33) node [anchor=north west][inner sep=0.75pt]   [align=left] {$\displaystyle y$};
\draw (297,150.33) node [anchor=north west][inner sep=0.75pt]  [font=\normalsize] [align=left] {$\displaystyle \theta $};
\draw (184.36,190.73) node [anchor=north west][inner sep=0.75pt]  [font=\normalsize] [align=left] {$\displaystyle {\displaystyle ( 0,0)}$};
\draw (148.28,187.82) node [anchor=north west][inner sep=0.75pt]   [align=left] {$\displaystyle R$};
\draw (280,84.33) node [anchor=north west][inner sep=0.75pt]  [font=\normalsize] [align=left] {$\displaystyle \mathbf{u}( \theta ) =\begin{bmatrix}
cos\theta \\
sin\theta 
\end{bmatrix}$};
\draw (253.5,202.28) node [anchor=north west][inner sep=0.75pt]  [font=\normalsize] [align=left] {$\displaystyle \mathbf{p}_{1} ,\ \phi _{1}$};
\draw (137.24,265.53) node [anchor=north west][inner sep=0.75pt]  [font=\normalsize] [align=left] {Ant 4};
\draw (71.5,189.44) node [anchor=north west][inner sep=0.75pt]  [font=\normalsize] [align=left] {Ant 3};
\draw (137.5,95.44) node [anchor=north west][inner sep=0.75pt]  [font=\normalsize] [align=left] {Ant 2};
\draw (253.5,189.28) node [anchor=north west][inner sep=0.75pt]  [font=\normalsize] [align=left] {Ant 1};
\draw (185.5,91.28) node [anchor=north west][inner sep=0.75pt]  [font=\normalsize] [align=left] {$\displaystyle \mathbf{p}_{2} ,\ \phi _{2}$};
\draw (60.5,203.28) node [anchor=north west][inner sep=0.75pt]  [font=\normalsize] [align=left] {$\displaystyle \mathbf{p}_{3} ,\ \phi _{3}$};
\draw (187.4,260.53) node [anchor=north west][inner sep=0.75pt]  [font=\normalsize] [align=left] {$\displaystyle \mathbf{p}_{4} ,\ \phi _{4}$};

\end{tikzpicture}
}

\begin{figure}[!t]
\centering
\subfloat[4-UCA spatial coordinate and wavefront projection]{%
  \resizebox{0.7\columnwidth}{!}{\usebox{\figonea}}%
  \label{fig:1a}%
}
\\
\subfloat[Mainlobe orientation vectors]{%
 \includegraphics[width=0.42\columnwidth]{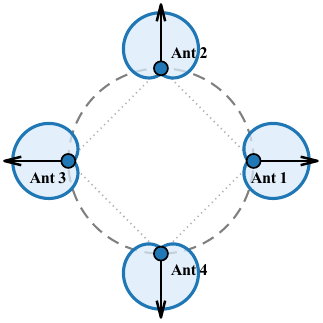}
  \label{fig:1b}%
}
\hspace{0.05\columnwidth}
\subfloat[Cardioid radiation pattern]{%
  \includegraphics[width=0.42\columnwidth]{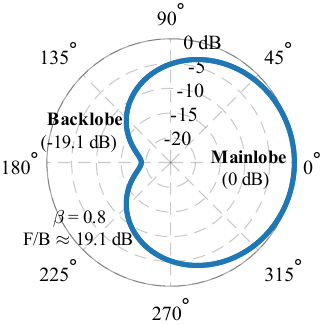}
  \label{fig:1c}%
}
\hfill

\caption{(a) 4-UCA spatial coordinate configuration and incoming wavefront projection, (b) radial mainlobe orientation vector distribution, and (c) first-order cardioid directivity model.}
\label{fig:system_geometry}
\end{figure}

This section establishes a generalized array manifold model coupling antenna spatial distribution with element radiation patterns. As shown in Fig.~\ref{fig:1a}, consider a 4-element uniform circular array (4-UCA) with radius $R$ on the 2D horizontal plane. Taking the array center as the origin, the coordinates of the $m$-th antenna element are $\mathbf{p}_m = [R \cos \phi_m, \, R \sin \phi_m]^T$, where the polar angle is $\phi_m = (m-1)\pi/2$ ($m=1,\dots,4$). For an incident planar wave from azimuth $\theta$ with unit directional $\mathbf{u}(\theta) = [\cos\theta, \, \sin\theta]^T$, the path-difference phase delay of the $m$-th element referenced to the origin is given by:

\begin{equation}\label{equ:2}
v_m(\theta) = e^{-j \frac{2\pi}{\lambda} \mathbf{p}_m^T \mathbf{u}(\theta)} = e^{-j \frac{2\pi R}{\lambda} \cos(\theta - \phi_m)},
\end{equation}
where $\lambda = c / f_c$ is the carrier wavelength.


As shown in Fig.~\ref{fig:1b}, we assume that the mainlobe of each antenna element points radially outward, with normal vector $\mathbf{n}_m = [\cos \phi_m, \, \sin \phi_m]^T$. Although \cite{Jackson2015} established the array response vector by incorporating element radiation patterns into UCA manifolds, its idealized cardioid pattern formulation $(1+\cos\theta)^m$ neglects rearward radiation. In practice, finite-sized ground planes and heat sinks inevitably induce rearward leakage, leading to a finite front-to-back ratio (FBR)~\cite{Jackson2015, Biguesh2009}. To model realistic radiation, we adopt a more generalized parameterized pattern formulation, as shown in Fig.~\ref{fig:1c}~\cite{Gazzah2018}:
\begin{equation}\label{equ:3}
g_m(\theta) = 1 + \beta (\mathbf{n}_m^T \mathbf{u}(\theta)) = 1 + \beta \cos(\theta - \phi_m),
\end{equation}
where $\beta \in [0, 1)$ is the sensor directivity parameter~\cite{Gazzah2018}. By tuning the scalar parameter $\beta$, the pattern can smoothly transition from an omnidirectional response to varying degrees of directivity, governing the achievable front-to-back ratio. Setting $\beta = 0.8$ as in~\cite{Gazzah2018} yields a nominal $\text{FBR} = 20\log_{10}[(1+\beta)/(1-\beta)]\approx19.1\text{ dB}$, representing a representative directional antenna response for the numerical analysis. In practical deployments, $\mathbf{g}(\theta)$ can also be derived directly from measured data obtained in the anechoic chamber. 

Based on \cite{Jackson2015}, by taking the Hadamard product of the phase delay vector $\mathbf{v}(\theta) = [v_1(\theta), \dots, v_4(\theta)]^T$ and the pattern vector $\mathbf{g}(\theta) = [g_1(\theta), \dots, g_4(\theta)]^T$, the final generalized array manifold vector is obtained as
\begin{equation}\label{equ:4}
\mathbf{a}(\theta) = \mathbf{g}(\theta) \odot \mathbf{v}(\theta).
\end{equation}

\subsection{ULA vs. UCA Directivity Effects}
In a 2D space, we first consider an ideal 8-element uniform circular array (8-UCA) with a half-wavelength adjacent chord spacing $0.5\lambda$ as an alias-free reference baseline; geometric calculations show that its array radius is $R \approx 0.65\lambda$. Constrained by the four RF chains on commercial Wi-Fi, deploying a 4-UCA across the same aperture $R$ expands the adjacent antenna spacing to a sparse $0.92\lambda$. To quantitatively evaluate the resulting undersampling grating lobes, a baseline 4-ULA is also configured with the identical sparse spacing of $0.92\lambda$. For an observed CSI vector $\mathbf{h}$, the normalized spatial power spectrum of matched-filtering is evaluated as 
\begin{equation}\label{equ:5}
P(\theta) = \frac{|\mathbf{a}^H(\theta) \mathbf{h}|^2}{\max_{\theta'} |\mathbf{a}^H(\theta') \mathbf{h}|^2}.
\end{equation}

\subsubsection{4-ULA}
\begin{figure}[!t]
\centering
\subfloat[Forward $\theta = 70^\circ$]{%
  \includegraphics[width=0.34\columnwidth]{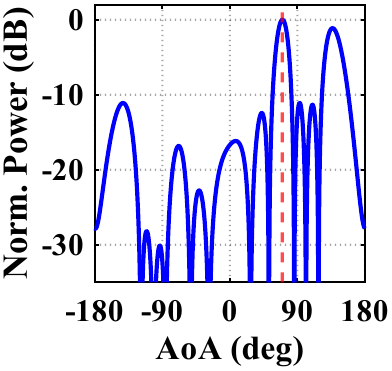}
  \label{fig:2a}%
}
\hfill
\subfloat[Rearward $\theta = -70^\circ$]{%
  \includegraphics[width=0.33\columnwidth]{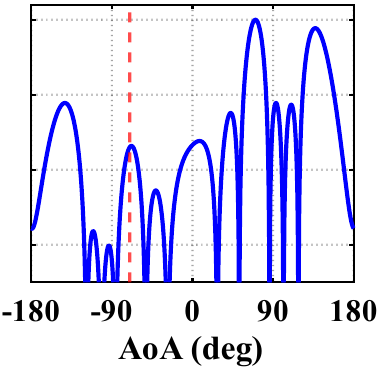}
  \label{fig:2b}%
}
\hfill
\subfloat[Transition $\theta = 35^\circ$]{%
  \includegraphics[width=0.33\columnwidth]{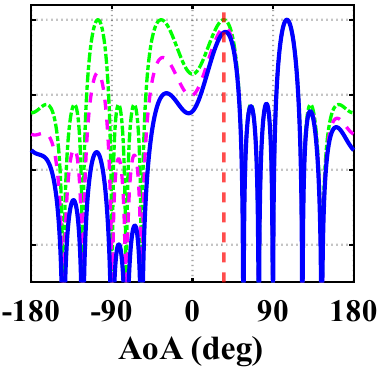}
  \label{fig:2c}%
}
\caption{Directional sparse 4-ULA: (a) unsuppressed grating lobes, (b) rearward mirror ambiguity, and (c) envelope roll-off across varying directivity factors: $\beta=0$ (green dash-dotted), $\beta=0.4$ (magenta dashed), and $\beta=0.8$ (blue solid).}
\label{fig:ula_degradation}
\end{figure}

All array elements are aligned along the $x$-axis in the $0^\circ$ direction, with their mainlobes synchronously pointing toward the positive $90^\circ$ broadside normal direction. For an arbitrary incident angle $\theta$, since the radiation patterns across all antennas are identical, the directivity vector degenerates into a common-mode scalar $\mathbf{g}(\theta) = g(\theta)\mathbf{1}_4$, simplifying the array manifold to a fixed envelope modulation $\mathbf{a}(\theta) \propto g(\theta)\mathbf{v}(\theta)$. As shown in Fig.~\ref{fig:ula_degradation}, this envelope induces fixed-pattern distortions across the spatial spectrum: for forward incidence at $70^\circ$, structural grating lobes remain unsuppressed, e.g., at $130^\circ$; for backward incidence at $-70^\circ$, the true target falls into a deep gain minimum, while the forward mirror grating lobes are amplified. In the endfire transition region $\theta = 35^\circ$, the steep pattern envelope roll-off suppresses the mainlobe even below the forward broadside grating lobe, as shown in Fig.~\ref{fig:2c}. Therefore, the element directivity fails to eliminate the grating lobes of the ULA. It shrinks the distortion-free sensing FoV to a narrow forward sector far less than $180^\circ$.

\subsubsection{4-UCA} 
\begin{figure}[!t]
\centering
\subfloat[Adaptive mainlobe tracking]{%
  \includegraphics[width=0.5\columnwidth]{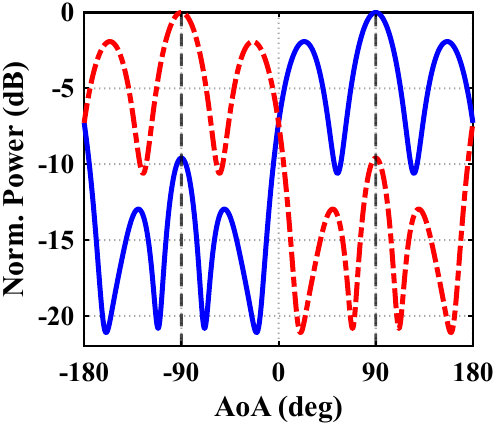}
  \label{fig:3a}%
}
\hfil
\subfloat[Tripartite pattern distortion]{%
  \includegraphics[width=0.49\columnwidth]{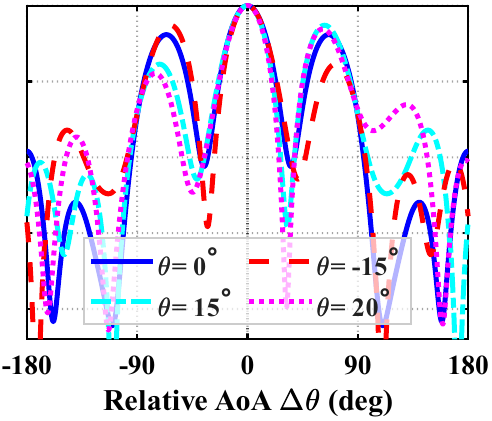}
  \label{fig:3b}%
}
\caption{Directional sparse 4-UCA: (a) mainlobe tracking and rearward attenuation under $\theta = +90^\circ$ (blue solid) and $\theta = -90^\circ$ (red dash-dotted), and (b) relative angular spectrums overlay showing tripartite pattern distortion.}
\label{fig:3}
\end{figure}

For the 4-UCA topology shown in Fig.~\ref{fig:1a},  the distinct mainlobe orientations of individual antennas introduce differential weighting to the phase delay vector $\mathbf{v}(\theta)$. Antennas facing the incident wave maintain high gain while rearward antennas experience attenuation. As shown in Fig.~\ref{fig:3}, this spatial filtering effect suppresses rearward grating lobes by $10\text{--}15\text{ dB}$ across the full $360^\circ$ azimuth plane. Furthermore, the antennas facing the target yield a persistent tripartite beam pattern, comprising the central mainlobe and two adjacent sidelobes. Its central mainlobe consistently locks onto the true AoA, endowing the directional 4-UCA with localization and tracking potential.

However, directivity cannot eliminate the inherent spatial undersampling defect. In Fig.~\ref{fig:3b}, spatial spectra under various incident angles are mapped onto a unified relative coordinate system centered at the true incident angle. The tripartite pattern exhibits significant morphological instability: at $0^\circ$, it retains a standard profile with sidelobes $2\text{ dB}$ below the mainlobe; at transition angles such as $\pm 15^\circ$, broken topological symmetry reduces the spectral margin between one adjacent sidelobe and the mainlobe to just $1\text{ dB}$; at singular angles like $20^\circ$, the tripartite pattern morphology becomes severely distorted. These fragile characteristics make AoA estimation and continuous tracking highly susceptible to outliers under noise fluctuations, regardless of whether maximum peak searching or tripartite pattern matching is employed.

\section{Beam-Switched UCA Architecture}
To overcome the sensing bottleneck of the directional 4-UCA topology, this section presents an angularly staggered beam-switched UCA architecture, detailing its hardware and the associated consensus spectrum synthesis framework.

\subsection{Angularly Staggered Beam-Switched UCA}
\begin{figure}[!t]
\centering
\includegraphics[width=0.85\columnwidth]{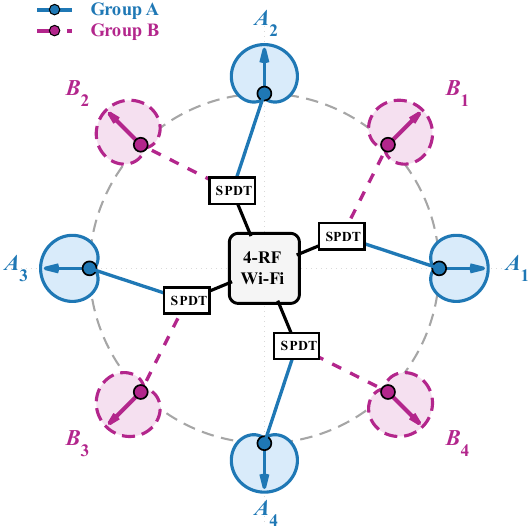}
\caption{Proposed beam-switched UCA architecture with 8 antennas and 4 single-pole double-throw (SPDT).}
\label{fig:4}
\end{figure}

 As shown in  Fig.~\ref{fig:4}, eight directional antennas are deployed across the same aperture radius $R \approx 0.65\lambda$ as aforementioned, and partitioned into two complementary 4-UCA groups angularly staggered by $45^\circ$. Group A inherits the geometry and radiation pattern in  Fig.~\ref{fig:1c}, with array manifold vector $\mathbf{a}_A(\theta) \in \mathbb{C}^{4 \times 1}$ according to (\ref{equ:4}); Group B is rotated by $45^\circ$ with polar coordinates $\phi_m^B = (2m-1)\pi/4 \in \{45^\circ, 135^\circ, 225^\circ, 315^\circ\}$ and manifold vector $\mathbf{a}_B(\theta) \in \mathbb{C}^{4 \times 1}$. Groups A and B are time-division multiplexed into the four RF chains of the Wi-Fi chip via four single-pole double-throw (SPDT) switches.

Groups A and B alternate on a packet-burst basis, balancing sensing continuity and communication throughput: inter-packet switching avoids disrupting intra-packet Wi-Fi demodulation, while the buffered CSI frames provide complementary spatial information. According to (\ref{equ:1}), when switched to Group A, the system updates the observed CSI vector $\mathbf{h}_A \in \mathbb{C}^{4 \times 1}$ of Group A in real time, while Group B retains the historical data from the last frame before switching. In contrast, the system updates $\mathbf{h}_B \in \mathbb{C}^{4 \times 1}$ in real time, while Group A retains the historical data. Through this alternating buffering, we acquire two complementary  spatial observations without increasing the number of RF transceiver chains.

\subsection{Complementary Consensus Spectrum Synthesis}
\begin{figure}[!t]
\centering
\subfloat[Consensus spectrum synthesis]{%
  \includegraphics[width=0.5\columnwidth]{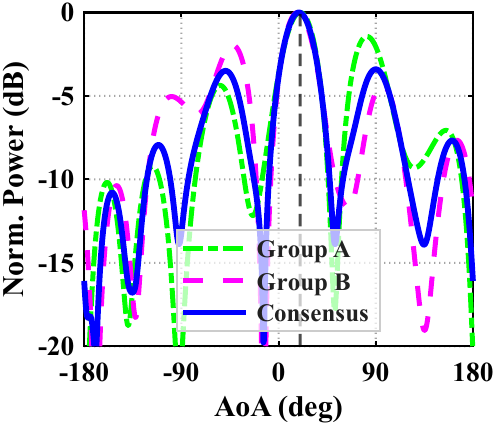}
  \label{fig:5a}%
}
\hfil
\subfloat[Full-azimuth relative overlay]{%
  \includegraphics[width=0.49\columnwidth]{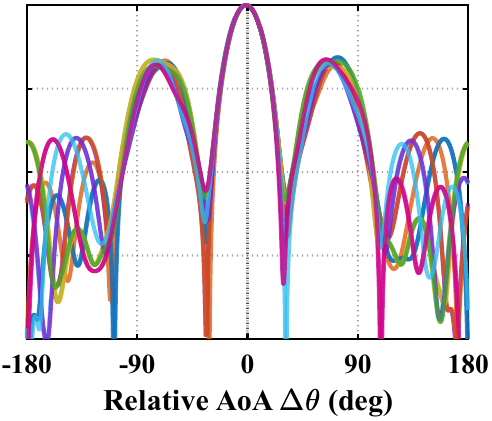}
  \label{fig:5b}%
}
\caption{Beam-switched UCA consensus spectrum: (a) consensus spatial spectrum synthesis at the angle $\theta = 20^\circ$, and (b) full-azimuth relative angular overlay for consensus spatial spectrum at $\theta = \pm 10^\circ, \pm 20^\circ, \pm 45^\circ,\pm 70^\circ,\pm 90^\circ,\pm 120^\circ,\pm 170^\circ$.}
\label{fig:5}
\end{figure}

The core rationale of this architecture lies in utilizing the $45^\circ$ spatial interleaving to break grating lobe ambiguities, ensuring that the sidelobe peaks of one group encounter spatial attenuation in the complementary group. After calculating the normalized spatial power spectra $P_A(\theta)$ and $P_B(\theta)$ of both groups according to (\ref{equ:5}), the joint spatial spectrum is synthesized using a nonlinear geometric mean consensus operator

\begin{equation}\label{equ:6}
P_{\text{syn}}(\theta) = \sqrt{P_A(\theta) \cdot P_B(\theta)}.
\end{equation}
This synthesis does not require strict phase coherence, but rather a relaxed scene coherence across the measurements. In the true direction of arrival, the mainlobes of both groups align coherently to maintain maximum gain. In contrast, in the sidelobe regions, the staggered peaks and troughs are multiplicatively suppressed by the deep opposing nulls. As shown in Fig.~\ref{fig:5a}, even at the typical singular angle $\theta = 20^\circ$ where the spectrum profile severely degrades, after consensus synthesis between Group A (green dashed line) and Group B (pink dashed line), the distorted spectrum is reshaped into a well-defined tripartite beam pattern (blue solid line). Comparing Fig.~\ref{fig:3} with Fig.~\ref{fig:5b}, the initially irregular and dispersed distortion envelopes are constrained into a consistent tripartite pattern across the full $360^\circ$ range of incidence, with the two sidelobes consistently remaining 3--4 dB below the mainlobe. This not only restores a robust decision margin, but also ensures omnidirectional geometric consistency, thereby mitigating tracking outliers induced by noise.

\
\section{Numerical Validation}

\begin{figure}[!t]
\centering
\subfloat[Directional 4-UCA]{%
  \includegraphics[width=0.445\columnwidth]{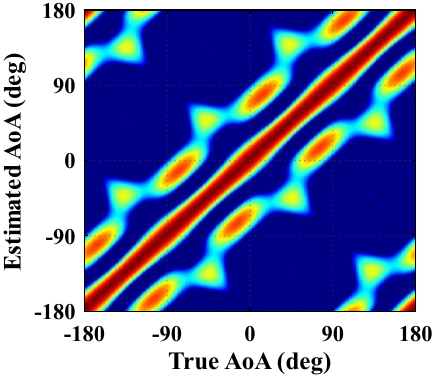}%
  \label{fig:6a}%
}
\hfill
\subfloat[Beam-switched UCA]{%
  \includegraphics[width=0.49\columnwidth]{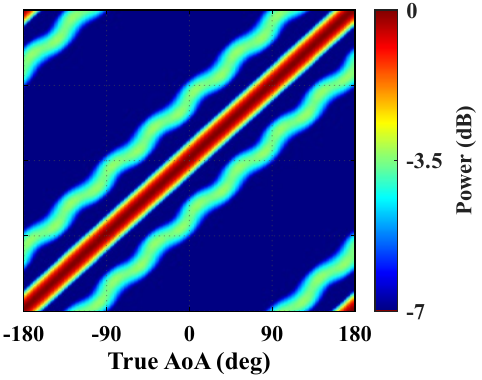}%
  \label{fig:6b}%
}
\\[-1.5ex] 
\subfloat[Omnidirectional 8-UCA]{%
  \includegraphics[width=0.445\columnwidth]{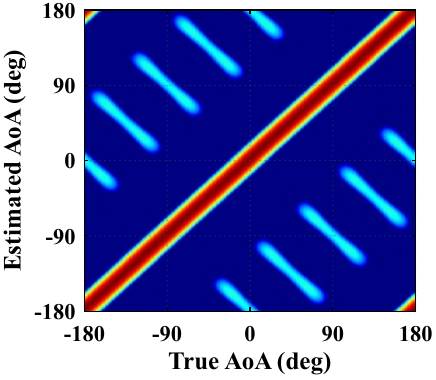}%
  \label{fig:6c}%
}
\hfill
\subfloat[Directional 8-UCA]{%
  \includegraphics[width=0.49\columnwidth]{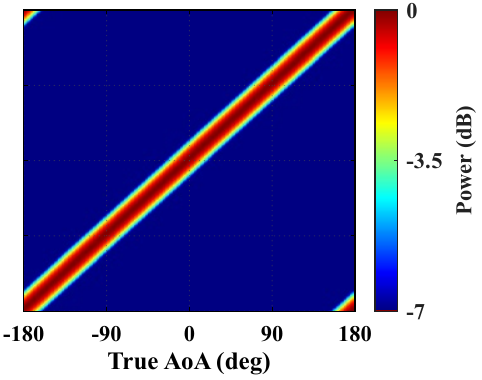}%
  \label{fig:6d}%
}
\caption{Spatial spectra across all AoAs: (a) directional 4-UCA, (b) proposed beam-switched UCA, (c) omnidirectional 8-UCA, and (d) directional 8-UCA.}
\label{fig:6}
\end{figure}

Simulations are conducted in the 5.745 GHz Wi-Fi band, adopting the array architecture configurations defined in the preceding sections. To isolate intrinsic array characteristics from complex multipath propagation, we adopt a line-of-sight (LOS) single-source benchmark under additive white Gaussian noise (AWGN) to evaluate the fundamental unambiguous angle-measuring limits of the arrays. Based on single-snapshot, single-carrier observations, the simulation performs matched filtering and peak searching across the full spatial grid at a $0.5^\circ$ resolution. The evaluation focuses on comparing the proposed 4-channel beam-switched UCA against the theoretical limits of the half-wavelength 8-UCA. An estimate deviating by more than $15^\circ$ from the ground truth exceeds the synthesized mainlobe coverage and is thus classified as an outlier.

Fig.~\ref{fig:6} compares the spatial spectra of the four architectures across all AoAs. Taking advantage of the identical aperture, all four configurations maintain consistent mainlobe widths. Although the beam-switched UCA and the omnidirectional 8-UCA exhibit similar grating lobe suppression capabilities, they display distinctly different patterns: the beam-switched UCA exhibits continuous sidelobes co-directional with angular variations, whereas the omnidirectional 8-UCA excites multiple segmented backlobes with reverse trajectories.

\begin{figure}[!t]
\centering
\subfloat[Directional 4-UCA]{%
  \includegraphics[width=0.445\columnwidth]{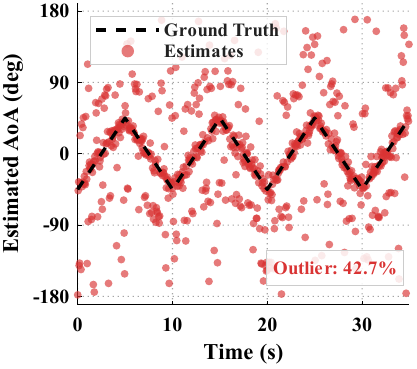}
  \label{fig:7a}%
}
\hspace{0.05\columnwidth}
\subfloat[Beam-switched UCA]{%
  \includegraphics[width=0.37\columnwidth]{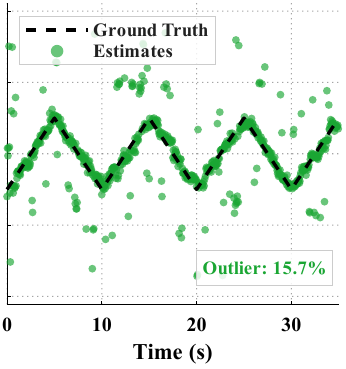}
  \label{fig:7b}%
}
\\[1.5ex]
\subfloat[Omnidirectional 8-UCA]{%
  \includegraphics[width=0.445\columnwidth]{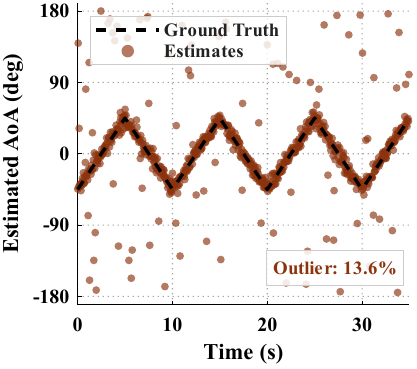}
  \label{fig:7c}%
}
\hspace{0.05\columnwidth}
\subfloat[Directional 8-UCA]{%
  \includegraphics[width=0.37\columnwidth]{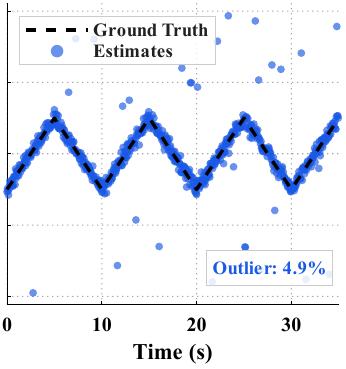}
  \label{fig:7d}%
}
\caption{Dynamic tracking trajectories ($\text{SNR} = 0\text{ dB}$): (a) 4-UCA, (b) beam-switched UCA, (c) omnidirectional 8-UCA, and (d) directional 8-UCA.}
\label{fig:7}
\end{figure}

Fig.~\ref{fig:7} illustrates the impact of these distinct patterns on dynamic tracking performance. Under $\text{SNR} = 0\text{ dB}$, we evaluate the tracking of a single target reciprocating within the $[-45^\circ, 45^\circ]$ range at a 20~Hz CSI frame rate. The directional 8-UCA achieves the highest accuracy, while the beam-switched UCA and the omnidirectional 8-UCA show similar accuracy levels. Furthermore, outliers of the omnidirectional 8-UCA are highly scattered and random, whereas the outliers of the beam-switched UCA are distributed along parallel sidelobes aligned with the motion, restricting errors within a bounded interval.

\begin{figure}[!t]
\centering
\includegraphics[width=0.8\columnwidth]{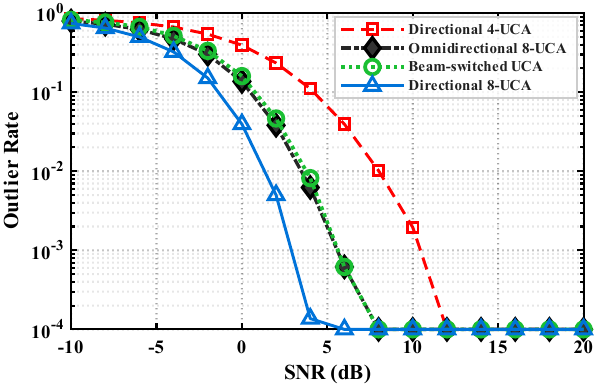}
\caption{Outlier rate versus SNR across all spatial directions.}
\label{fig:8}
\end{figure}

Fig.~\ref{fig:8} shows the curves of outlier rate versus SNR across the full space. For each angle, 400 Monte Carlo trials with added noise are conducted for stationary targets across $360^\circ$ with  $5^\circ$ intervals. The results show that the curve of the proposed beam-switched UCA closely matches the performance of the omnidirectional 8-UCA, with the outlier rate dropping to zero at $\text{SNR} = 7.5\text{ dB}$, showing an overall performance second only to the directional 8-UCA as the idealized 8-channel benchmark. This is mainly attributed to dual mechanisms: first, the spatial complementarity of alternating groups effectively suppresses grating lobes; second, the consensus synthesis of the two groups smooths out random noise fluctuations. The proposed architecture achieves the noise robustness of the half-wavelength omnidirectional 8-UCA merely by relying on a sparse UCA with four channels.

\section{Conclusion}
This paper investigates the AoA estimation mechanism and grating lobe suppression methods for 4-channel, large-aperture sparse arrays. Theoretical analysis reveals that element directivity acts as a spatial filter to attenuate rearward grating lobes in sparse UCAs, yet sparse 4-UCAs still suffer from severe sidelobe competition. To resolve this, the proposed angularly staggered beam-switched UCA architecture employs two complementary bases for nonlinear consensus synthesis, effectively suppressing grating lobes without precise coherent stitching or complex virtual-array algorithms. Numerical evaluations demonstrate that this architecture achieves localization and tracking performance comparable to an 8-channel omnidirectional reference array. At $\text{SNR} = 0\text{ dB}$, it reduces the dynamic tracking outlier rate from 42.7\% for the directional 4-UCA to 15.7\% and constrains outliers within sidelobe regions parallel to the motion trajectory, thus confirming the feasibility of omnidirectional passive sensing under the hardware constraints of 4-channel commercial Wi-Fi.


\section*{Acknowledgment}

The work is funded by Marie Skłodowska-Curie Actions (MSCA) Doctoral Network SMARTTEST under GA no. 101167834.



%

\end{document}